\documentclass[letterpaper, 10 pt, conference]{ieeeconf}
\IEEEoverridecommandlockouts    
\usepackage{graphics} % for pdf, bitmapped graphics files
\usepackage{epsfig} % for postscript graphics files
\usepackage{times} % assumes new font selection scheme installed
\usepackage{amsmath} % assumes amsmath package installed
\usepackage{amssymb}  % assumes amsmath package installed

\usepackage{subfigure}
\usepackage{mathtools}
\usepackage{color}
\usepackage{cite}
\usepackage{comment}
\usepackage{balance}
\usepackage[table]{xcolor}
\usepackage{multirow}
\usepackage{hyperref}
\hypersetup{
    colorlinks=true,
    linkcolor=black,
    citecolor=green,
    urlcolor=blue,
}
\usepackage{blindtext}
\usepackage{bm}
\usepackage{algorithm}
\usepackage{algpseudocode}
\usepackage[official]{eurosym}
\usepackage{multirow} % Required for multirow feature
\usepackage{booktabs} % For prettier tables

\makeatletter
\def\endthebibliography{%
  \def\@noitemerr{\@latex@warning{Empty `thebibliography' environment}}%
  \endlist
}
\makeatother

\algrenewcommand\algorithmicindent{1.0em}
\algnewcommand\algorithmicswitch{\textbf{switch}}
\algnewcommand\algorithmiccase{\textbf{case}}
\algnewcommand\algorithmicassert{\texttt{assert}}
\algnewcommand\Assert[1]{\State \algorithmicassert(#1)}%
\algdef{SE}[SWITCH]{Switch}{EndSwitch}[1]{\algorithmicswitch\ #1\ \algorithmicdo}{\algorithmicend\ \algorithmicswitch}%
\algdef{SE}[CASE]{Case}{EndCase}[1]{\algorithmiccase\ #1}{\algorithmicend\ \algorithmiccase}%
\algtext*{EndSwitch}%
\algtext*{EndCase}%

\makeatletter
\newcommand{\algmargin}{\the\ALG@thistlm}
\makeatother
\newlength{\forwidth}
\algdef{SE}[parFOR]{parFor}{EndparFor}[1]
  {\parbox[t]{\dimexpr\linewidth-\algmargin}{%
     \hangindent\forwidth\strut\algorithmicfor\ #1\ \algorithmicdo\strut}}{\algorithmicend\ \algorithmicfor}%
\newlength{\forif}
\algnewcommand{\parState}[1]{\State%
  \parbox[t]{\dimexpr\linewidth-\algmargin}{\strut #1\strut}}

\title{\LARGE \bf
Small-Scale Testbed for Evaluating \\ C-V2X Applications on 5G Cellular Networks
}

\author{Kaj Munhoz Arfvidsson$^{*,1}$, Kleio Fragkedaki$^{*,1}$, Frank J. Jiang$^{*,1}$, Vandana Narri$^{1,2}$, \\ Hans-Cristian Lindh$^3$, Karl H. Johansson$^1$, Jonas Mårtensson$^1$%
\thanks{
    $^*$Indicates equal contribution}%
\thanks{
    This work was partially supported by the Swedish Innovation agency (Vinnova), under grant 2021-02555 Future 5G Ride, within the Strategic Vehicle Research and Innovation program (FFI); the Wallenberg Artificial Intelligence, Autonomous Systems, and Software Program (WASP) funded by the Knut and Alice Wallenberg Foundation; the Swedish Research Council Distinguished Professor Grant 2017-01078; and the Knut and Alice Wallenberg Foundation Wallenberg Scholar Grant.}% <-this % stops a space
\thanks{
    $^{1}$ are with the Division of Decision and Control Systems, EECS, KTH Royal Institute of Technology, Malvinas v{\"a}g 10, 10044 Stockholm, Sweden, {\tt\small \{kajarf, kfrag, frankji, narri, kallej, jonas1\}@kth.se}. They are also affiliated with the Integrated Transport Research Lab and Digital Futures.}%`
\thanks{
    $^{2}$ is with Research and Development, Scania CV AB, 151 87 Södertälje, Sweden, {\tt\small vandana.narri@scania.com}.}%
\thanks{
    $^{3}$ is a Project Manager with Standards and Technology, Ericsson AB, Torshamnsgatan 21, 164 40 Kista, Stockholms län, Sweden, {\tt\small hans-christian.xh.lindh@ericsson.com}.}%
}

\begin{document}

\maketitle
\thispagestyle{empty}
\pagestyle{empty}

%%%%%%%%%%%%%%%%%%%%%%%%%%%%%%%%%%%%%%%%%%%%%%%%%%%%%%%%%%%%%%%%%%%%%%%%%%%%%%%%
\begin{abstract}
In this work, we present a small-scale testbed for evaluating the real-life performance of cellular V2X (C-V2X) applications on 5G cellular networks. Despite the growing interest and rapid technology development for V2X applications, researchers still struggle to prototype V2X applications with real wireless networks, hardware, and software in the loop in a controlled environment. To help alleviate this challenge, we present a testbed designed to accelerate development and evaluation of C-V2X applications on 5G cellular networks. By including a small-scale vehicle platform into the testbed design, we significantly reduce the time and effort required to test new C-V2X applications on 5G cellular networks. With a focus around the integration of small-scale vehicle platforms, we detail the design decisions behind the full software and hardware setup of commonly needed intelligent transport system agents (e.g. sensors, servers, vehicles). Moreover, to showcase the testbed's capability to produce industrially-relevant, real world performance evaluations, we present an evaluation of a simple test case inspired from shared situational awareness. Finally, we discuss the upcoming use of the testbed for evaluating 5G cellular network-based shared situational awareness and other C-V2X applications. 
\end{abstract}

%%%%%%%%%%%%%%%%%%%%%%%%%%%%%%%%%%%%%%%%%%%%%%%%%%%%%%%%%%%%%%%%%%%%%%%%%%%%%%%%
\section{Introduction}\label{sec:intro}
%% INTRODUCTION

Over the recent years, the interest in exploring the use of vehicle-to-everything (V2X) communication has grown immensely. Due to factors like unprotected road-users and occlusions, there are complex traffic scenarios that are difficult for automated vehicles to resolve individually. To safely and efficiently handle these scenarios, automated vehicles need to communicate with each other, infrastructure, and central servers. This has led to the development of various Intelligent Transport Systems (ITS) that utilize wireless technologies including different varieties of dedicated short-range, direct communication, 4G, and 5G for V2X communication to enhance traffic safety and improve road throughput~\cite{Soto2022}. Vehicles that utilize V2X communication to enhance their automation are referred to as connected and automated vehicles (CAV). Moreover, throughout this work, we refer to the collection of ITS services that leverage V2X communication as V2X applications. 

\begin{figure}[t]
    \centering
    \vspace{0.25cm}
    \includegraphics[width=0.98\linewidth]{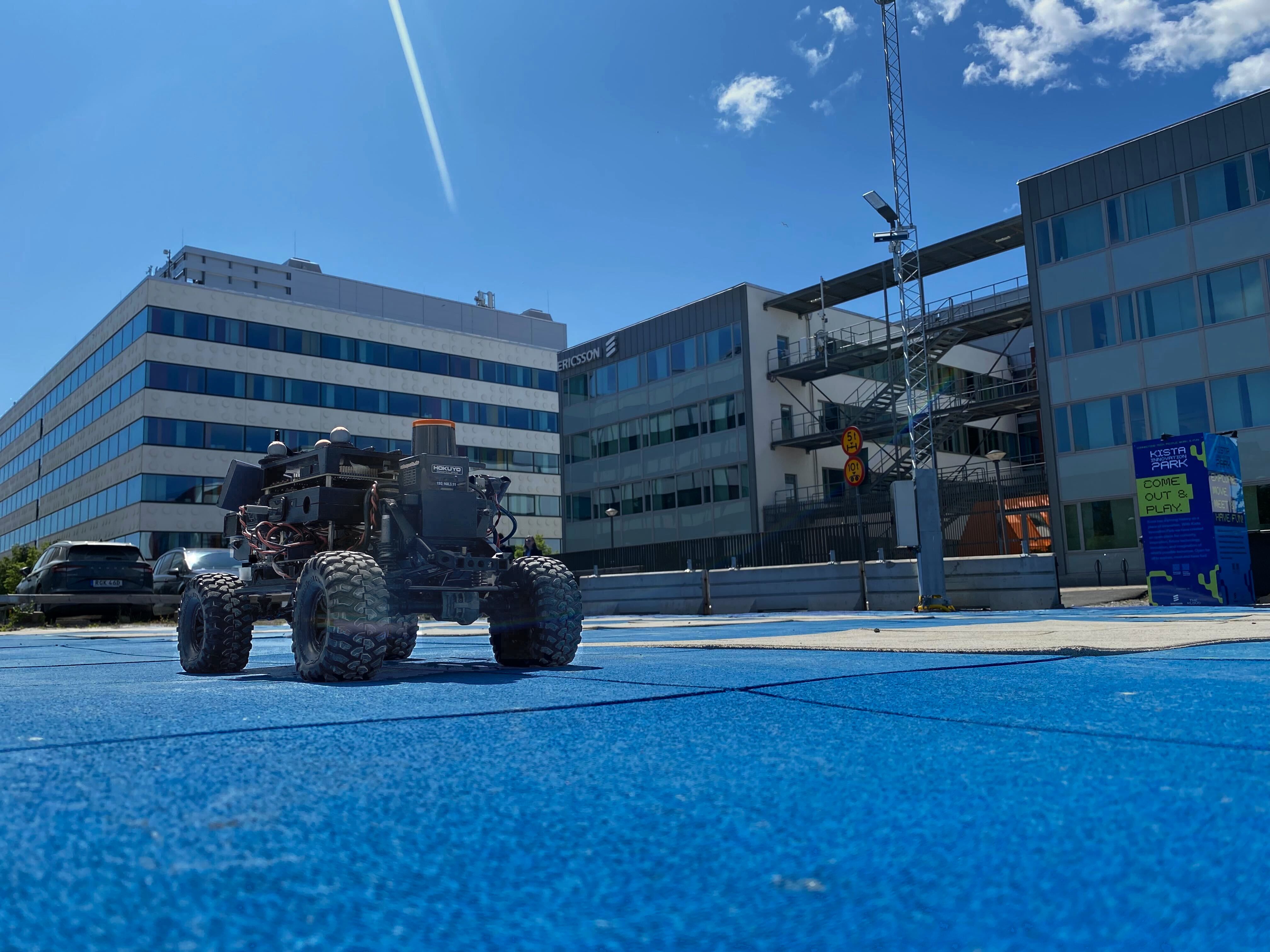}
    \caption{A snapshot of the 1/10th-scale SVEA platform driving in the presented 5G-based C-V2X testbed in the Kista Innovation Park. More details about the SVEA platform at [\url{https://svea.eecs.kth.se}]}
    \label{fig:svea}
\end{figure}

Already today, there are a variety of V2X applications that have been developed using various wireless technologies. These V2X applications are typically add-on services to the core functionality of CAVs that leverage communicated data to implement functionalities like beyond line-of-sight collision warnings, intersection movement assistance, or left turn assistance~\cite{Amjad2018, Miucic2018}. V2X applications are developed to further enhance safety and efficiency in scenarios where the sensor suite onboard the CAV is unable to perceive critically important information. One example of this is shared situational awareness, where sensors, servers, and vehicles share their perception data with each other using V2X direct communication to enhance each other's situational awareness~\cite{Narri_Alanwar_2023}. V2X applications are predominantly developed around one of two main V2X communication standards: ITS-G5~\cite{etsi_its_g5} and C-V2X~\cite{3GPP}. ITS-G5 is based on Wi-Fi communication on the 5.9 GHz band. C-V2X is an umbrella technology which encapsulates all 3GPP V2X technologies, including both direct communication and cellular network communications~\cite{5GAutomativeAssociation5GAA}. While there are some preliminary studies performed in~\cite{Tahir2020} on C-V2X using 5G cellular networks, many of the studies conducted so far are developed around using 4G. We attribute this to the current difficulty around prototyping advanced C-V2X applications in a controlled environment with a real 5G network, hardware, and software in the loop.

To help alleviate this difficulty, we design and develop a testbed specifically for evaluating the real-life performance of advanced C-V2X applications on 5G cellular networks. More specifically, we design a testbed to enable low-cost experimentation on advanced C-V2X applications with the ability to precisely evaluate the 5G cellular network, hardware, and software's performance. To make experimentation cheaper, we build the testbed around the use of a 1/10th-scale CAV platform. Although 1/10th-scale CAVs do not capture the full dynamics of full-scale CAVs, they do provide the ability to cheaply evaluate V2X applications with real network, hardware, and software in the loop~\cite{Jiang_Al-Janabi_Bolin_Johansson_Martensson_2022}. Additionally, 1/10th-scale CAVs offer the ability to do preliminary studies with motion, providing some initial insight into how the results translate to full-scale CAVs.

\subsection{Contribution}
In summary, the main contribution of this paper is the design of a testbed for rapidly evaluating the real-life performance of C-V2X applications on 5G cellular networks. Explicitly, our contributions are three-fold:
\begin{enumerate}
    \item we overview the design choices behind the network configuration, software, and hardware for supporting small-scale evaluations of the real-life performance of C-V2X applications on 5G cellular networks,
    \item we showcase the testbed's capability to measure the real-life performance of a simple, situational awareness-inspired, test case under the realistic conditions of: clock synchronization error, network traffic overloading, and base station handovers,
    \item we discuss the implications from the test case on the future evaluation of 5G-based C-V2X applications.
\end{enumerate}
Additionally, the software used for sending C-V2X data over the 5G cellular network is available publicly.\footnote{https://github.com/KTH-SML/cv2x-testbed}

\section{Motivation}\label{sec:overview}
% Preliminary Material
In this section, we elaborate on the motivation for developing the testbed presented in this paper.  Despite the significant effort and work that has previously been put into proposing V2X solutions, researchers still struggle to develop and evaluate the end-to-end performance of V2X applications with a real hardware in the loop. Given the relatively recent introduction of 5G cellular networks, this difficulty is especially apparent. To this end, the motivating challenges we address in this work are the following:
\begin{enumerate}
    \item how do we develop a modular testbed where 5G cellular network-based C-V2X applications can be implemented with less lead time for testing and evaluation?
    \item what developments are necessary to replicate scenarios with 1/10th-scale vehicles under realistic conditions?
\end{enumerate}
To address the first challenge, we go through the design decisions that make it easier to develop and test C-V2X applications on 5G cellular networks. These design decisions are carefully made while considering the trade-offs between development speed and resultant performance. Throughout the work, we note the trade-offs of our design decisions and mention alternatives that could improve performance with the trade-off of increasing implementation difficulty.

To address the second challenge, we implement a simplified test case inspired by the requirements of shared situational awareness. In this test case, we adapt the architecture in~\cite{Narri_Alanwar_2023} to better suit 5G cellular networks and implement a generic version of the messaging required to support these types of applications. We test this adapted architecture under the realistic conditions of clock synchronization error, network traffic overloading, and base station hand-over events. While we draw inspiration from the implementations done in~\cite{Narri_Alanwar_2023}, we have chosen not to implement the full shared situational awareness application so we can obtain results that are more relatable to other C-V2X applications. Moreover, since the implementation is done with 1/10th-scale vehicles, the available software and hardware that can be integrated greatly differs from full-scale vehicles. At 1/10th-scale, we need to work with hardware and software more typical in robotics and mobile device development.

\begin{figure*}[t]
    \centering
    \includegraphics[width=0.98\linewidth]{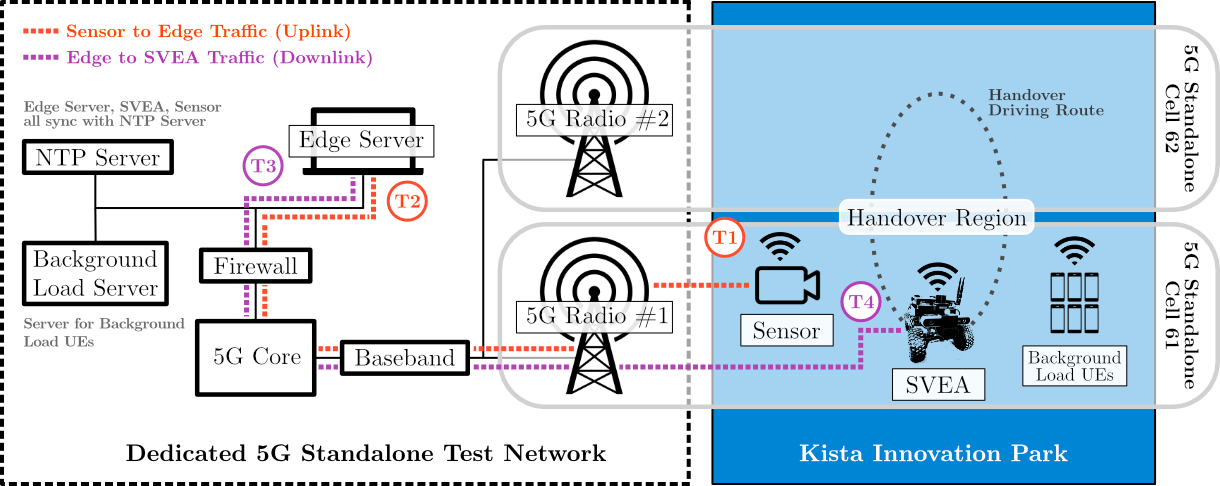}
    \caption{Architecture and Spatial Distribution of our 5G C-V2X Testbed System. There are three types of ITS agents: infrastructure sensor, edge server and CAVs (SVEA). To communicate, the sensor and CAVs have a wireless connection using a 5G cellular network with 20 MHz bandwidth. Each radio cell has ca. 40 Mbps capacity. T1, T2, T3, T4 are the transmission time stamps of the messages used in our evaluation. All agents rely on an NTP server for clock synchronization. The NTP server is reached using the same wired or wireless connection as is used for their main communication.}
    \label{fig:system}
\end{figure*}

\section{Testbed Design}\label{sec:design}
% Testbed Design

In this section, we present our testbed (illustrated in Fig.~\ref{fig:system}) for 5G-cellular network-based C-V2X applications and how it can evaluate end-to-end performance. In particular, we detail the important design decisions behind the dedicated 5G standalone test network, ITS agent implementations, messaging system, and clock synchronization software in our testbed. For simplicity, we refer to 5G-cellular network-based C-V2X as ``5G C-V2X'' for the remainder of the paper. 

\subsection{Dedicated 5G Standalone Test Network}

To fully evaluate a 5G C-V2X application, it is necessary to recreate real-life conditions that impact the cellular network. Central to our testbed is a dedicated 5G standalone test network that allows for custom network configuration. With it, we can precisely modify background load and enable features that protect traffic, ensuring a realistic setting for experiments. It is important to investigate the trade-offs related to network performance as it is not only influenced by the traffic volume but also by the number of users, the direction of the traffic, and the configuration of the network. Our testbed's network has a  20 MHz bandwidth and uses a Time Division Duplex (TDD) pattern \verb|DDDSU|, where \verb|D| represents a downlink slot, \verb|U| represents an uplink slot, and \verb|S| represents a special slot with 10 downlink symbols and 2 uplink symbols. Furthermore, a radio cell, in this network, has maximum capacity of ca. 40 Mbps. Lastly, in the context of our scenario, we utilize a specific QoS mechanism in the 5G radio access network called absolute priority scheduling. 

\subsection{ITS Agents}
For quick and low-cost evaluation of 5G C-V2X, the three agents (sensor, edge server, CAV/SVEA) shown in Fig.~\ref{fig:system} are actualized with small-scale platforms. They implement their 5G C-V2X applications using the Robot Operating System (ROS) \cite{ros}. The use of ROS yields significant development benefits, as it provides wide, open-source support for most of the sensor and message types that are required in C-V2X applications. For example, in~\cite{Kueppers2024}, authors contribute an open-source package containing ROS support for ETSI ITS messages for V2X communication. However, since ROS is a best-effort implementation, it is not the most suitable middleware for external communication between agents on the 5G cellular network. Thus, we only use ROS on the local networks of each ITS agent. External communication can instead be facilitated by messaging systems such as MQTT \cite{mqtt311}, NATS \cite{CNCF_Synadia} or Zenoh\cite{Zenoh}. These have a direct impact on the overall 5G C-V2X performance, and the choice often depends on various design aspects like application compatibility and preferred implementation language. In line with the objectives of this platform, to accelerate development and evaluation of 5G C-V2X applications, we use the Transmission Control Protocol (TCP)-based messaging system NATS with the NATS broker sitting on the edge server in Fig.~\ref{fig:system}. Since NATS supports distributed brokers, it enables high-performing implementations of edge architectures for V2X applications.

\subsubsection{\textbf{Sensor}}
The sensor in our 5G C-V2X testbed is designed to be application-agnostic. It represents any device that produces data within the network but does not consume it. This can characterize a wide range of different applications and concepts, such as local infrastructure sensors or bystander CAVs. For external communication, it is reasonable to assume that an infrastructure sensor has a wired connection. However, in certain use cases or scenarios, a sensor may require the flexibility of a wireless connection. For this reason, in this work, the sensor is communicating  with the other agents through the cellular 5G network. Ultimately, this leads to easier, more agile setups for outdoor experiments. However, if latency and reliability need to be minimized, a wired connection can be used instead.

\subsubsection{\textbf{Edge Server}}
The edge server has the role of a centralized computing unit that aggregates and processes information and may also engage in decision-making processes. In our testbed, it acts as a proxy between the infrastructure sensor and the CAVs. Any computation that should be done on the server can be emulated by delaying messages. Depending on the scenario and application, the server can be located locally or more centrally. By being physically close to its users, the server can deliver benefits such as reduced latency and higher bandwidth. This is enabled by the 5G network's ability to deploy applications near the ``edge'' of an access network. However, it may be tolerable in some applications to instead deploy in far-away data centers, which can be more convenient or provide access to more computational hardware. Since we are prioritizing latency and bandwidth in our applications, we deploy the server on the edge in our testbed setup.

\subsubsection{\textbf{1/10th-scale CAV (SVEA)}}
The vehicle has been developed as a stand-alone platform for evaluating 5G C-V2X use cases. Introduced in \cite{Jiang_Al-Janabi_Bolin_Johansson_Martensson_2022}, the SVEA is a 1/10th-scale automated vehicle that consists of the necessary components, both in hardware and software, to build an autonomous system. This enables the platform to evaluate a wide range of scaled-down traffic scenarios. Moreover, the platform can be used to collect and log vehicle state data, such as velocity and GPS position, even when driving manually. This makes the platform ideal for network-centric experiments that do not require any autonomy, but where mobility still plays an important role. In contrast to the sensor, in the test case we set up, the SVEA represents a user in the network that only consumes data.

\subsection{Messaging}
In our testbed, communication between the agents occurs in both uplink and downlink channels. As illustrated in Fig.~\ref{fig:system}, the infrastructure sensor sends messages to the edge server, which then distributes those messages to the vehicle. 
We name the connections after the primary direction of the application-critical traffic. That is, UL for uplink from sensor to server and DL for downlink from server to vehicle. However, due to TCP, which requires acknowledgments to be sent in the opposite direction, there also exist downlink traffic from UL and uplink from DL. 

The traffic is characterized by the transmitted messages. A message contains data of a certain size, and is sent at a certain rate. The contained data can be in any format, however, a message must include the timestamps, clock offsets, a sequence number, and a checksum to verify the message was sent correctly. This information will also later be used to determine latency. Before the sensor sends a message, it sets the $T_1$ timestamp from the system clock time. Once the message is received by the edge server, it updates the messages with a corresponding timestamp $T_2$. In the same way, when the edge server sends the message, and a vehicle receives it, then $T_3$ and $T_4$ are updated as depicted in Fig.~\ref{fig:system}.

The performance of 5G C-V2X applications can be evaluated with metrics such as latency, jitter, signal-to-noise ratio, packet drops and packet error rates. While the primary benefit of NATS is effortless deployment of edge architectures, since it is built on TCP, many of these performance metrics aggregates to latency due to TCP's order guarantees, error checks and retransmission algorithms. This simplifies the evaluation such that we can focus on measuring latency. For applications that need further latency minimization, technologies like MQTT can be used with UDP, which introduces the need to explicitly measure packet drops and packet error rates. However, by using MQTT with UDP, the implementation of edge architectures becomes more difficult since MQTT brokers cannot be directly distributed without additional implementations. As previously described, messages are sent in both UL and DL traffic. Therefore, we can also distinguish between $L_\text{UL}$ and $L_\text{DL}$ latencies. If all agents were perfectly synchronized, the latency would simply be the difference between the respective timestamps, e.g., $L_\text{UL} = T_2 - T_1$.

\subsection{Network Time Protocol for Synchronization} \label{sec:kpi}

Measuring latency between different agents is difficult due to synchronization errors. Consider $T_1$ and $T_2$ which are measured on different clocks. This becomes a major source of errors when considering how computer clocks can drift non-deterministically depending on voltage fluctuations, temperature changes, system load, and production quality. Consequently, taking the time difference $T_2 - T_1$ would not only measure the communication latency, but also include any synchronization error between the clocks. 

In our testbed all agents use Network Time Protocol (NTP) \cite{ntp}  to both correct their system clock and periodically query the NTP server for synchronization errors. By querying the NTP server during a test, we can correct for synchronization error in the latency measurement. As such, when setting timestamps in the messages, we also set a corresponding clock offset value. To account for the synchronization error, the latency becomes $L_\text{UL} = (T_2 + E_2) - (T_1 + E_1)$.

While alternative methods like GPS Clocks or Precision Time Protocol (PTP) \cite{ptp} can be utilized for more precise time synchronization, we have excluded them from our design due to hardware restrictions. Specifically, since server implementations are typically installed indoors, there is not always infrastructure available to provide servers access to a GPS receiver. Furthermore, we do not always have access to network infrastructures with hardware that has PTP support throughout. In the case of software-based PTP, the benefit over NTP is not as apparent when in a wireless network due to PTP being more sensitive to delay variations. Therefore, we choose to use the NTP client chrony\footnote{https://chrony-project.org/} for clock synchronization due to it being readily available on most operating systems.

\section{Testbed Evaluation}\label{sec:results}
In this section, we develop and evaluate a simple test case to assess the testbed's ability to support 5G C-V2X applications under various conditions. Specifically, we present the test case's latency under three distinct conditions that could impact the overall efficiency or reliability of 5G C-V2X applications. Our primary objective in this section is to show how the testbed can be used to evaluate the real-life performance of a 5G C-V2X application. To this end, we start by contextualizing our evaluation in a previously developed V2X application and use that application to determine a relevant architecture for us to test.

\begin{table}[b]
    \centering
    \renewcommand{\arraystretch}{1.1}
    \begin{tabular}{cclcc}
        \hline
        & \textbf{\begin{tabular}[c]{@{}c@{}}Network\\ Config.\end{tabular}} & \textbf{\begin{tabular}[c]{@{}c@{}}Background Load\\ Uplink -- Downlink\\ {[Mbps]}\end{tabular}} & \multicolumn{1}{c}{\textbf{\begin{tabular}[c]{@{}c@{}}Message\\ Size\\ {[kB]}\end{tabular}}} & \textbf{\begin{tabular}[c]{@{}c@{}}Message\\ Frequency\\ {[Hz]}\end{tabular}} \\ 
        \hline \hline
        \multirow{8}{*}{\rotatebox[origin=c]{90}{Nominal}} & \multirow{4}{*}{BL} & \multirow{2}{*}{No Load -- No Load}& 1 & 10 \\
        & & & 10 & 20 \\ 
        \cline{3-5} 
        & & \multirow{2}{*}{\begin{tabular}[c]{@{}c@{}}1 UEs x  5 -- 1 UEs x 110\end{tabular}} & 1 & 10 \\
        & & & 10 & 20 \\ 
        \cline{2-5} 
        & \multirow{4}{*}{AP} & \multirow{2}{*}{No Load -- No Load} & 1 & 10 \\
        & & & 10 & 20 \\ 
        \cline{3-5} 
        & & \multirow{2}{*}{\begin{tabular}[c]{@{}c@{}}1 UEs x 5 -- 1 UEs x 110\end{tabular}} & 1 & 10 \\
        & & & 10 & 20 \\ 
        \hline
        \multirow{4}{*}{\rotatebox[origin=c]{90}{Overload}} & \multirow{2}{*}{BL} & 1 UEs x 40 -- No Load & 10 & 20 \\
        & & 2 UEs x 40 -- No Load & 10 & 20 \\ 
        \cline{2-5} 
        % & & 0 UEs x 40 - 1 UEs x 300 & 10 & 20 \\
        % & & 0 UEs x 40 - 2 UEs x 300 & 10 & 20 \\ 
        % \cline{2-5} 
        & \multirow{2}{*}{AP} & 1 UEs x 40 -- No Load & 10 & 20 \\
        & & 2 UEs x 40 -- No Load & 10 & 20 \\ 
        % \cline{3-5} 
        % & & 0 UEs x 40 - 1 UEs x 300 & 10 & 20 \\
        % & & 0 UEs x 40 - 2 UEs x 300 & 10 & 20 \\ 
        \hline
        \rotatebox[origin=c]{90}{ Mobility } & BL & No Load -- No Load & 10 & 20 \\ 
        \hline
    \end{tabular}
    \caption{Combinations of network configurations, background load, message sizes, and frequencies in the presented test case.}
    \label{tab:combinations_experiments}
    \vspace{-2em}
\end{table}

\subsection{Test Case Context: Shared Situational Awareness}

\begin{figure}[t]
    \centering
    \includegraphics[width=0.8
\linewidth]{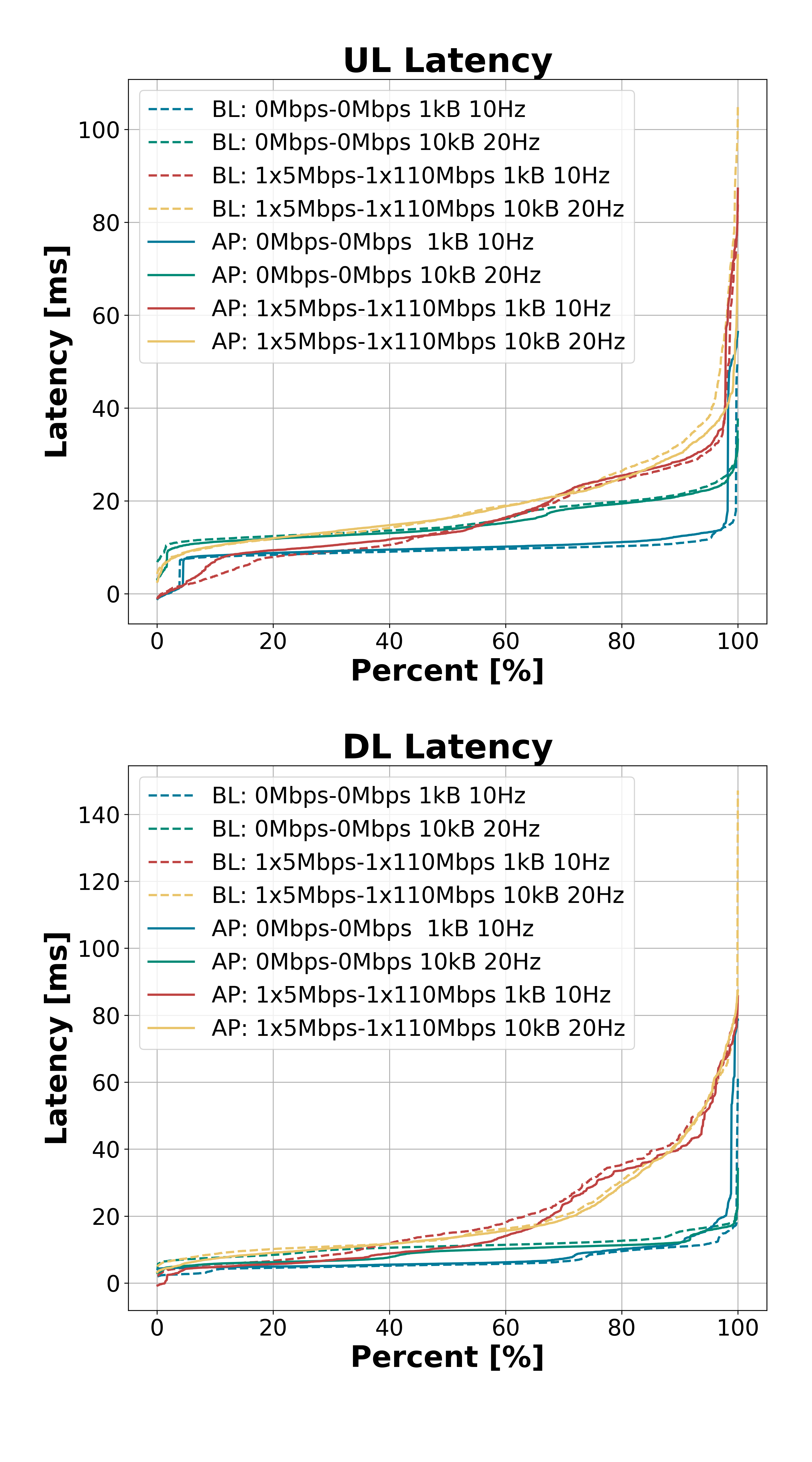}
    \caption{Cumulative Distribution Functions (CDFs) of latency for all the different combinations under nominal conditions.}
    \label{fig:nominal_condition_tests_cdfs}
\end{figure}

For our test case, we are inspired by~\cite{Narri_Alanwar_2023}, where the authors have implemented a shared situational awareness framework, with an objective to improve situational awareness of the ego-vehicle using V2X communication. In this cited paper, a simple urban scenario is considered, where the field-of-view of the ego-vehicle has an occlusion created by a parked vehicle. But this occluded region is visible for two other connected V2X units in the near surrounding. The connected V2X unit shares their perception data with the ego-vehicle.

The authors utilize the standardized  Collaborative Perception Message (CPM)~\cite{ETSI_TR_103_562}. In this standard, a general message structure consists of Management Container (MC), Station Data Container (SDC), Sensor Information Container (SIC), Perceived Object Container (POC), and Free Space Addendum (FSA). In~\cite{Narri_Alanwar_2023}, the communication of CPMs is carried out over an ITS-G5 access layer with a 10 MHz bandwidth. Each message is sent and received at a frequency of 10 Hz. The total payload of each message was 156 bytes and each message consists of MC, SIC, and POC. In each message, there were five POCs, which had a payload of 78 bytes. According to IEEE 802.11 standards, the ITS-G5 network has a limitation to send each message with a maximum payload of 1500 bytes. This means that if messages need to be sent with payloads more than 1500 bytes, the messages need to be sent in separate parts. Based on these values, we are able to determine some of the relevant message sizes and frequencies to include in our test case.

\begin{figure}[b]
    \centering
        \label{subfig:50percentload_nominal_packet_number}
        \includegraphics[width=0.95
            \linewidth]{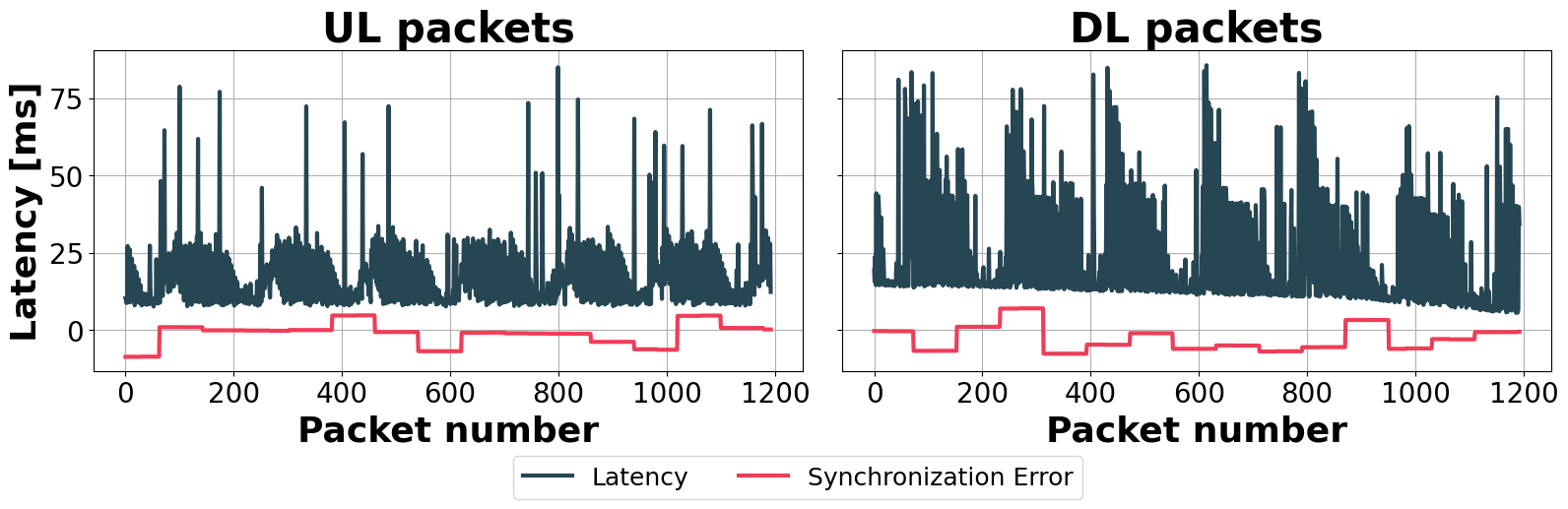}
    \caption{Latency of individual packets under background load. 
    This test was conducted with message size and frequency of 1 kB at 10 Hz, in BL network configuration, while there was 1 UEs x 5 Mbps uplink and 1 UEs x 110 Mbps downlink load. In red, we see the synchronization error between sensor-server and server-vehicle, respectively.}
    \label{fig:packet_number_latency_nominal}
\end{figure}

\begin{figure}[t]
    \centering
        \centering
        \includegraphics[width=.9\linewidth]{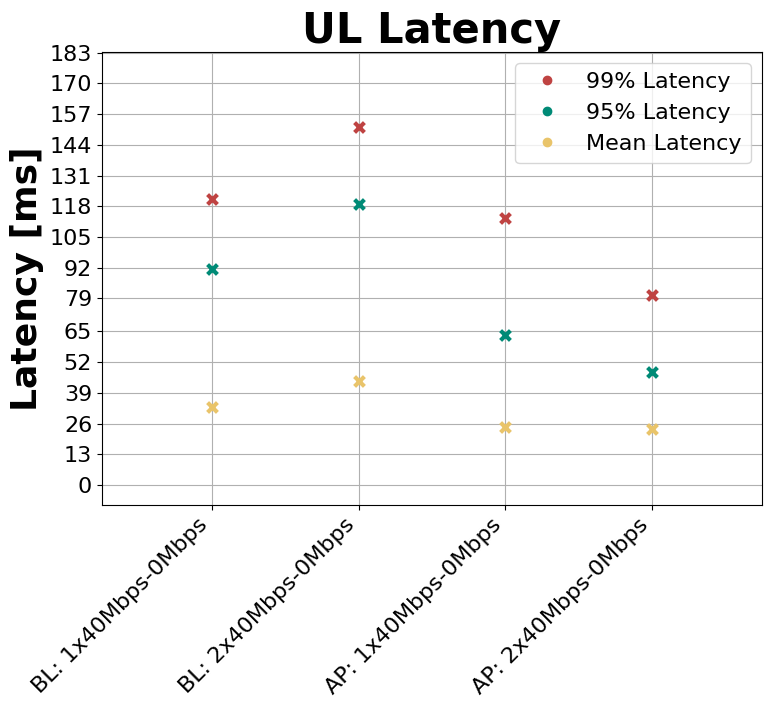}
        \caption{Comparison between different cases under overload conditions. The figure depicts 99\%, 95\%, and mean latency when overloading the network in uplink traffic.}
        \label{fig:overload_plots}
\end{figure}

In this work, we evaluate our testbed with a simple test case that evaluates the messaging capabilities within the 5G cellular network to develop a preliminary understanding of how a 5G C-V2X shared situational awareness system could be designed. Since 5G networks have different bandwidth alternatives and different features for prioritizing sessions in the network, we set up a test case where we evaluate the testbed's performance with varying network loads, message sizes, and message frequencies. Moreover, since safety-critical data is being communicated, it's of great interest to understand the effects of network traffic overloading and handovers on the messaging between agents. As the focus of the developed testbed is the facilitation of development and evaluation of advanced 5G C-V2X applications, we will not focus on comparing the performance of 5G C-V2X with ITS-G5-based V2X. Instead, through our test, we are interested in understanding how shared situational awareness would look as a 5G C-V2X application.

\begin{figure}[t]
    \centering
    
    % First Row
    \begin{subfigure}{}
        \centering
        \includegraphics[width=0.9\linewidth]{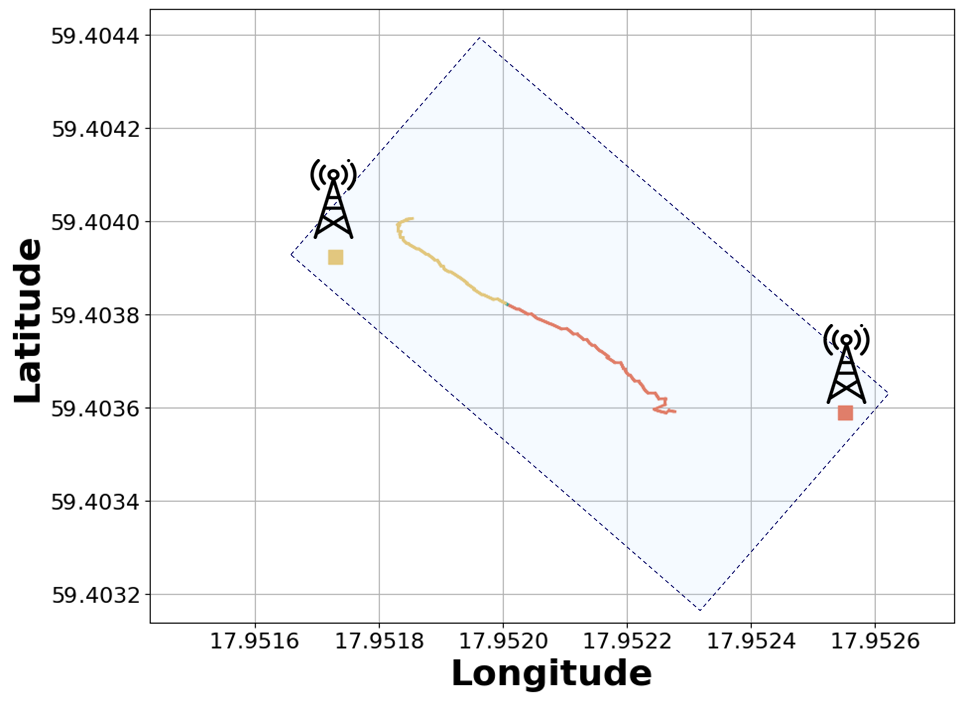}
    \label{subfig:milti-ho}
    \end{subfigure}%
    
    % Second Row
    \begin{subfigure}{}
        \centering
        \includegraphics[width=0.95\linewidth]{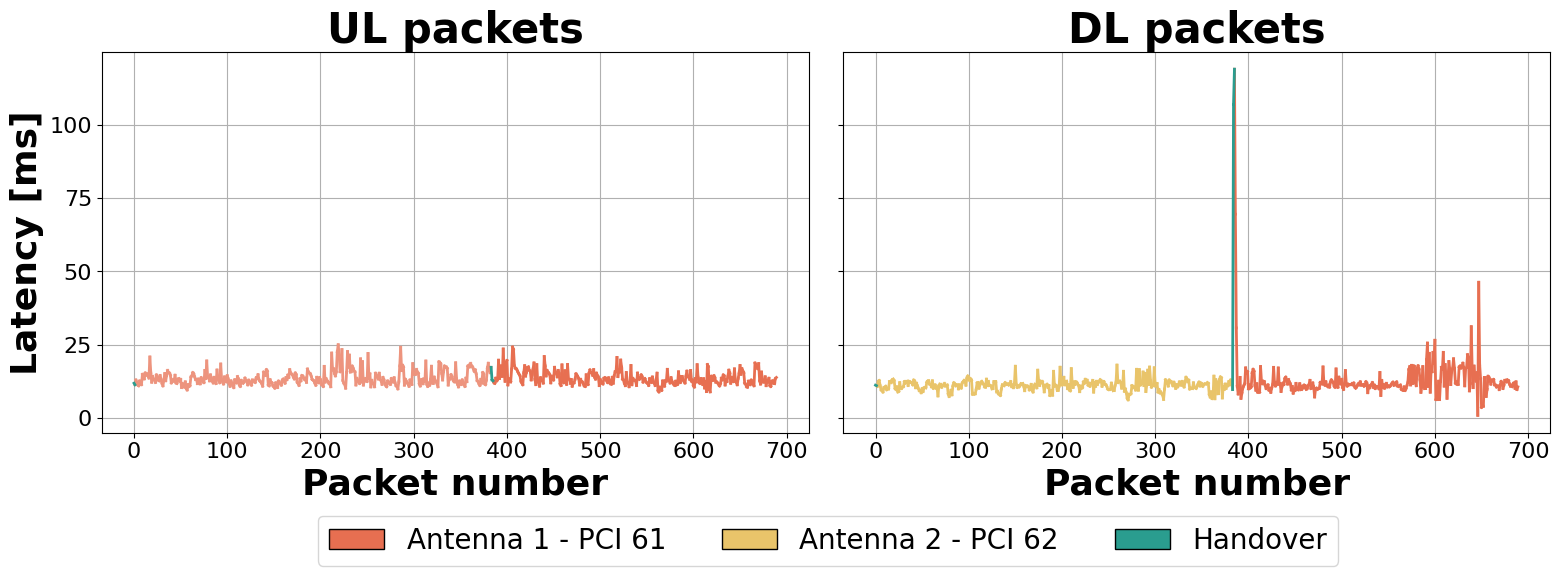}
        
        \label{subfig:packet_number_multi-ho}
    \end{subfigure}%
    
    \caption{Vehicle traveling along the handover route. The vehicle starts near Antenna 2 (yellow) and drives towards Antenna 1 (orange). During this route, DL traffic (server-vehicle) is logged and shown in the respective color. UL traffic (sensor-server), which always goes through Antenna 1, is also shown before (light orange) and after (dark orange) handover. Packets affected by the handover are shown in green.}
    \label{fig:multi-ho_example}
\end{figure}

\subsection{Test Case Setup}

Our testbed is in the Kista Innovation Park (KIP), an arena that provides a controlled environment for conducting research and development of C-V2X applications on a dedicated 5G standalone test network. As illustrated in Fig.~\ref{fig:system}, KIP is equipped with two radio stations, allowing for the study of handover events in a controlled manner. This, combined with the ability to freely configure its network, positions KIP as a suitable arena to explore real-world complexities, yet under controlled and manageable circumstances.

Based on the urban scenario introduced for shared situational awareness, for our test case, we employ an infrastructure sensor, an edge server, and a 1/10th-scale CAV as the primary agents. By setting up this type of test case, we can measure the latency of safety critical perception data when sent from a sensor to a CAV through an edge server over a 5G cellular network, as shown in Fig.~\ref{fig:system}. We can also compare the effect of different network configurations. In this work, we show this capability using a QoS mechanism in the 5G network named absolute priority scheduling that aims to protect our safety-critical traffic. We compare the absolute priority (AP) configuration with a baseline (BL) that uses default scheduling.  

In our implementation, we send messages containing dummy data instead of real perception messages to more easily control certain traffic characteristics such as message size and frequency. The different combinations of network settings, background loads, message sizes, and frequencies applied in our experiments are summarized in Table~\ref{tab:combinations_experiments}. With these, we show how flexible our testbed is in terms of how quickly one can develop V2X scenarios and start testing different designs of 5G C-V2X applications. By exposing these options, we can create test cases relatable to existing ITS-G5 based standards, i.e. 1 kB messages at 10 Hz, then continue to investigate other test cases that take better advantage of the features in 5G networks. We note that these test cases are not exhaustive and in the future it'd be interesting to continue to test shared situational awareness under other scenarios. Next, we describe the specific network conditions we considered in this work.

\subsubsection{\textbf{Nominal Conditions}}
In the nominal conditions, we consider a static 1/10th-scale CAV that is receiving messages of a particular size and at a particular frequency. At the same time, there is background load in the network generated by regular User Equipment (UE) using iPerf \cite{iperf}, and this load is carried over User Datagram Protocol (UDP). Although the application data is carried over TCP through NATS, the background load has to flood the network, making UDP a more suitable transport protocol. For the 5G network, that has 20 MHz bandwidth, we consider 1 UE loading 5 Mbps in the uplink direction and 1 UE loading 110 Mbps in the downlink direction to be within nominal conditions with respect to the network's capacity.

\subsubsection{\textbf{Overload Conditions}}
In the overload conditions, we are also analyzing conditions when there is background load exceeding the network's capacity, which we refer to as overloading conditions. Since networks are typically not configured for symmetric resource allocation in uplink and downlink traffic, it can be further valuable to test background load in each direction separately. To assess the potential benefits of prioritized network traffic on 5G C-V2X performance, we test the network configurations AP and BL when there is asymmetric background load.

\subsubsection{\textbf{Mobility Conditions}}
In the mobility conditions, due to the motion of the vehicle, handover events are triggered when the vehicle transitions between the two radio cells, as illustrated in Fig.~\ref{fig:system}. These events can increase latency, which is why they are an important aspect to consider during mobility conditions.

\subsection{Test Case under Nominal Conditions}

Under nominal conditions, in a static vehicle agent setting, we examine our test case with varying message types, background load, and network configurations, as detailed in Table~\ref{tab:combinations_experiments}. 
%Using the testbed, someone can further investigate the effect that different traffic characteristics and network configurations have on latency.
Using Cumulative Distribution Functions (CDFs) of latency,
% that is measured according to Section \ref{sec:kpi},
we can compare the impact of different traffic and network attributes with respect to the given requirements of a 5G C-V2X application. For example, a requirement could be that 95\% of the latency distribution must be less than 40 ms. Fig.~\ref{fig:nominal_condition_tests_cdfs} shows such CDFs for the test cases under nominal conditions. As expected, there is no significant difference between BL and AP network configurations, since the benefit of absolute priority scheduling is not apparent with such low load and just one background UE in the cell.

In addition to Fig.~\ref{fig:nominal_condition_tests_cdfs}, it can also be useful to view the latencies of individual packets. Since our testbed logs each message, it is possible to perform such analyses, as depicted in Fig.~\ref{fig:packet_number_latency_nominal}. It is worth noting that this particular test was affected by synchronization errors due to drifting system clocks. Depending on what is being tested, this might not be a significant problem. However, this issue demonstrates the importance of careful consideration of, and planning for, time synchronization. 

\subsection{Test Case under Overload Conditions}
Overload conditions, as the name conveys, describe the case when there is background traffic exceeding the network's capacity. In Fig.~\ref{fig:overload_plots}, we present the 99\%, 95\%, and mean latency values of the UL traffic when there is extreme background load only in the uplink direction. Through these measurements, we see that the network was able to protect the UL traffic under overload conditions with the help of the enabled 5G features in AP.

\subsection{Test Case under Mobility Conditions}
Evaluating network behavior in the context of a moving vehicle is particularly interesting due to the handover events that can occur. The network configuration, message size, and message frequency of this experiment are described in Table~\ref{tab:combinations_experiments}, while no additional features were added to help lower the handover latency. Fig.~\ref{fig:multi-ho_example} displays the GPS coordinates and packet latency following the handover route in Fig.~\ref{fig:system}. Specifically, this allows us to know where and when these handovers occur, along with the latency at these times. By following this route in a controlled environment, we can repeat handover events to evaluate their consistency and the 5G C-V2X application performance when these transitions occur. 

\section{Discussion}\label{sec:disc}
% Discussion

The performed tests, inspired from shared situational awareness, show that the presented testbed implementation stands as a fast and effective development platform. Moreover, it indicates that small-scale testbeds are suitable for evaluating advanced 5G C-V2X applications. For example, from the test cases under nominal conditions, we are able to track individual messages and measure the latency when they are of varying sizes and frequencies. From the test cases under overload conditions, we are able to observe the testbed's ability to stress test an implemented 5G C-V2X application when there are additional users in the same network. Finally, from the test cases under mobility conditions, we are able to see the effects of handovers, providing us a controlled way to evaluate these events in relation to the 5G C-V2X application. For future work, we are implementing the full 5G C-V2X shared situational awareness application along with an intelligent intersection application, similar to~\cite{MunhozArfvidsson2024}, and will evaluate both of their real-life performances. In parallel to this implementation, we will also continue to investigate and implement new features, such as features limiting handover interruption time, that are being specified in 3GPP to enhance the network's ability to support 5G C-V2X applications
\balance

%%%%%%%%%%%%%%%%%%%%%%%%%%%%%%%%%%%%%%%%%%%%%%%%%%%%%%%%%%%%%%%%%%%%%%%%%%%%%%%%

\section*{ACKNOWLEDGMENT}
The authors would like to thank Mustafa Al-Janabi for software contributions to the messaging system used in this work.

%%%%%%%%%%%%%%%%%%%%%%%%%%%%%%%%%%%%%%%%%%%%%%%%%%%%%%%%%%%%%%%%%%%%%%%%%%%%%%%%
\bibliographystyle{IEEEtran}
\bibliography{references}

\begin{thebibliography}{10}
\providecommand{\url}[1]{#1}
\csname url@samestyle\endcsname
\providecommand{\newblock}{\relax}
\providecommand{\bibinfo}[2]{#2}
\providecommand{\BIBentrySTDinterwordspacing}{\spaceskip=0pt\relax}
\providecommand{\BIBentryALTinterwordstretchfactor}{4}
\providecommand{\BIBentryALTinterwordspacing}{\spaceskip=\fontdimen2\font plus
\BIBentryALTinterwordstretchfactor\fontdimen3\font minus \fontdimen4\font\relax}
\providecommand{\BIBforeignlanguage}[2]{{%
\expandafter\ifx\csname l@#1\endcsname\relax
\typeout{** WARNING: IEEEtran.bst: No hyphenation pattern has been}%
\typeout{** loaded for the language `#1'. Using the pattern for}%
\typeout{** the default language instead.}%
\else
\language=\csname l@#1\endcsname
\fi
#2}}
\providecommand{\BIBdecl}{\relax}
\BIBdecl

\bibitem{Soto2022}
I.~Soto, M.~Calderon, O.~Amador, and M.~Urue{\~{n}}a, ``{A survey on road safety and traffic efficiency vehicular applications based on C-V2X technologies},'' \emph{Vehicular Communications}, vol.~33, p. 100428, 2022.

\bibitem{Amjad2018}
Z.~Amjad, A.~Sikora, B.~Hilt, and J.-P. Lauffenburger, ``{Low Latency V2X Applications and Network Requirements: Performance Evaluation},'' in \emph{2018 IEEE Intelligent Vehicles Symposium (IV)}, pp. 220--225.

\bibitem{Miucic2018}
R.~Miucic, A.~Sheikh, Z.~Medenica, and R.~Kunde, ``{V2X Applications Using Collaborative Perception},'' in \emph{2018 IEEE 88th Vehicular Technology Conference (VTC-Fall)}, pp. 1--6.

\bibitem{Narri_Alanwar_2023}
\BIBentryALTinterwordspacing
V.~Narri, A.~Alanwar, J.~M{\aa}rtensson, C.~Nor{\'{e}}n, and K.~H. Johansson, ``{Shared Situational Awareness with V2X Communication and Set-membership Estimation},'' no. arXiv:2302.05224, 2023. [Online]. Available: \url{http://arxiv.org/abs/2302.05224}
\BIBentrySTDinterwordspacing

\bibitem{etsi_its_g5}
\BIBentryALTinterwordspacing
{European Telecommunications Standards Institute}, ``{Intelligent {Transport} {Systems} ({ITS}); {ITS-G5} {Access} layer specification for Intelligent Transport Systems operating in the {5$\sim$GHz} frequency band}.'' [Online]. Available: \url{https://www.etsi.org/}
\BIBentrySTDinterwordspacing

\bibitem{3GPP}
\BIBentryALTinterwordspacing
3GPP, ``{The 3rd generation partnership project}.'' [Online]. Available: \url{https://www.3gpp.org}
\BIBentrySTDinterwordspacing

\bibitem{5GAutomativeAssociation5GAA}
\BIBentryALTinterwordspacing
{5G Automative Association (5GAA)}, ``{C-V2X explained}.'' [Online]. Available: \url{https://5gaa.org/c-v2x-explained/}
\BIBentrySTDinterwordspacing

\bibitem{Tahir2020}
M.~N. Tahir, T.~Sukuvaara, and M.~Katz, ``{Vehicular Networking: ITS-G5 vs 5G Performance Evaluation using Road Weather Information},'' in \emph{2020 International Conference on Software, Telecommunications and Computer Networks (SoftCOM)}, pp. 1--6.

\bibitem{Jiang_Al-Janabi_Bolin_Johansson_Martensson_2022}
F.~J. Jiang, M.~Al-Janabi, T.~Bolin, K.~H. Johansson, and J.~Mårtensson, ``{SVEA: an experimental testbed for evaluating V2X use-cases},'' in \emph{2022 IEEE 25th International Conference on Intelligent Transportation Systems (ITSC)}.\hskip 1em plus 0.5em minus 0.4em\relax Macau, China: IEEE, pp. 3484--3489.

\bibitem{ros}
\BIBentryALTinterwordspacing
{Stanford Artificial Intelligence Laboratory et al.}, ``Robotic operating system.'' [Online]. Available: \url{https://www.ros.org}
\BIBentrySTDinterwordspacing

\bibitem{Kueppers2024}
G.~Kueppers, J.-P. Busch, L.~Reiher, and L.~Eckstein, ``V2aix: A multi-modal real-world dataset of etsi its v2x messages in public road traffic,'' \emph{arXiv preprint arXiv:2403.10221}, 2024.

\bibitem{mqtt311}
{OASIS MQTT Technical Committee}, ``{MQTT Version 3.1.1 Protocol Specification},'' OASIS, Tech. Rep., 2014, available: http://docs.oasis-open.org/mqtt/mqtt/v3.1.1/os/mqtt-v3.1.1-os.html.

\bibitem{CNCF_Synadia}
\BIBentryALTinterwordspacing
{Cloud Native Computing Foundation and Synadia}, ``{NATS}.'' [Online]. Available: \url{https://nats.io}
\BIBentrySTDinterwordspacing

\bibitem{Zenoh}
\BIBentryALTinterwordspacing
{ZettaScale Technology}, ``Zenoh.'' [Online]. Available: \url{https://zenoh.io}
\BIBentrySTDinterwordspacing

\bibitem{ntp}
\BIBentryALTinterwordspacing
J.~Martin, J.~Burbank, W.~Kasch, and P.~D.~L. Mills, ``{Network Time Protocol Version 4: Protocol and Algorithms Specification},'' RFC 5905, Jun. 2010. [Online]. Available: \url{https://www.rfc-editor.org/info/rfc5905}
\BIBentrySTDinterwordspacing

\bibitem{ptp}
``{IEEE Standard for a Precision Clock Synchronization Protocol for Networked Measurement and Control Systems},'' \emph{{IEEE Std 1588-2019 (Revision of IEEE Std 1588-2008)}}, pp. 1--499, 2020.

\bibitem{ETSI_TR_103_562}
\BIBentryALTinterwordspacing
ETSI, \emph{{Intelligent Transport Systems (ITS); Vehicular Communications; Basic Set of Applications; Analysis of the Collective Perception Service (CPS); Release 2}}, etsi tr 103 562 v2.1.1 (2019-12)~ed.\hskip 1em plus 0.5em minus 0.4em\relax The European Telecommunications Standards Institute, 2019. [Online]. Available: \url{https://www.etsi.org/deliver/etsi_tr/103500_103599/103562/02.01.01_60/tr_103562v020101p.pdf}
\BIBentrySTDinterwordspacing

\bibitem{iperf}
\BIBentryALTinterwordspacing
L.~ESnet, ``iperf3.'' [Online]. Available: \url{http://software.es.net/iperf/}
\BIBentrySTDinterwordspacing

\bibitem{MunhozArfvidsson2024}
K.~M. Arfvidsson, F.~J. Jiang, K.~H. Johansson, and J.~Mårtensson, ``{Ensuring Safety at Intelligent Intersections: Temporal Logic Meets Reachability Analysis},'' in \emph{2024 IEEE Intelligent Vehicles Symposium (IV)}.

\end{thebibliography}

%%%%%%%%%%%%%%%%%%%%%%%%%%%%%%%%%%%%%%%%%%%%%%%%%%%%%%%%%%%%%%%%%%%%%%%%%%%%%%%%

\end{document}